# ATOMS AND FORCES OF INTERACTION BETWEEN ELEMENTARY PARTICLES IN THE EXPANDING UNIVERSE

M.V. Gorbatenko, V.P. Neznamov[1]

RFNC-VNIIEF, 37 Mira Pr., Sarov, 607188 Russia

Abstract

The earlier developed algorithm for constructing a self-conjugate Hamiltonian in the $\eta$ representation for Dirac particles interacting with a general gravitational field is extended to the case of electromagnetic fields. This Hamiltonian is applied to the case when the gravitational field describes the spatially flat Friedmann model, and the electromagnetic field is the Coulomb potential extended to the case of this model.

The analysis of atomic systems and electromagnetic forces of interaction under the conditions of spatially flat expansion of the universe has demonstrated that the system of atomic levels does not change with cosmological time. Spectral lines of atoms in the spatially flat Friedmann model are identical at different points of cosmological time. In this case the redshift is stipulated entirely by the growth of wavelength of photons at movement in the expending universe. At the same time force of interaction between elementary particles can change with expansion of the universe.

---

[1] e-mail: neznamov@vniief.ru

# 1. INTRODUCTION

In [1] we developed an algorithm for constructing a self-conjugate Hamiltonian in the $\eta$ representation for Dirac particles interacting with a general gravitational field. Using this algorithm, in the same study we developed self-conjugate Hamiltonians for spin ½ particles in the expanding universe described by Friedmann solutions. In the current study, we use obtained Hamiltonian to investigate a question of influence of the gravitational field on the states of Dirac particles in coupled systems, such as hydrogen-like atoms. We consider the spatially flat Friedmann model, which is known (see e.g. [2]) to quite adequately describe the early and current evolution of the universe.

For this model with the metric

$$ds^2 = -c^2 dt^2 + b^2(t)\left[ dx^2 + dy^2 + dz^2 \right], \tag{1}$$

the Hamiltonian in the $\eta$ representation for a Dirac particle with mass $m$ within the notation of Ref. [1] takes the following form:

$$\mathrm{H}_\eta = im\gamma_{\underline{0}} - \frac{i}{b(t)} \gamma_{\underline{0}} \gamma_{\underline{k}} \frac{\partial}{\partial x^k}. \tag{2}$$

In a quasi-stationary approximation for the cosmological time $t$ the energy operator can be written as

$$E = \sqrt{H_\eta^2} = \sqrt{m^2 + \frac{\mathbf{p}^2}{b^2(t)}}. \tag{3}$$

In expressions (2), (3) and below, the system of units is $\hbar = c = 1$; $p^k = -i\frac{\partial}{\partial x^k}$ are components of the momentum of the Dirac particle.

It follows from the expressions that the physical momentum of Dirac particles is given by the quantity

$$\mathbf{p}_{phys} = \frac{\mathbf{p}}{b(t)}. \tag{4}$$

In the expanding universe, the quantity $b(t)$ grows with time, and in the accompanying cosmological frame of reference, physical momentums $\mathbf{p}_{phys}(t)$ decrease with time. Spin ½ Dirac particles are gradually freezing in. The situation is absolutely identical to the analysis in [1] of the behavior of photons and massive spinless particles in the expanding universe. Originally relativistic Dirac particles with mass $m$ become non-relativistic on the large cosmological time.

It is of interest to look at the coupled atomic states of a Dirac particle in the spatially flat Friedmann model.

In Sec. 2 we establish the form of Hamiltonian (2) for the case of interaction with an electromagnetic field. In Sec. 3 we obtain solutions to the Maxwell equations in the spatially flat Friedmann model. In Sec. 4 we consider atomic systems and electromagnetic forces of interaction for the spatially flat expansion of the universe.

As a result, we can provide strong validation of the fact that the system of atomic levels does not vary with cosmological time. And, on the contrary, the forces of interaction between elementary particles vary as the universe expands.

Conclusion provides a general discussion of the results of this study.

## 2. HAMILTONIAN OF A DIRAC PARTICLE INTERACTING WITH A GRAVIATIONAL AND ELECTROMAGNETIC FIELD

In accordance with [1], the resulting Hamiltonian $H_\eta$ in the $\eta$ representation for a Dirac particle in a gravitational field takes the following form:

$$H_\eta = -\frac{im}{\left(-g^{00}\right)}\tilde{\gamma}^0 + \frac{i}{\left(-g^{00}\right)}\tilde{\gamma}^0\tilde{\gamma}^k\frac{\partial}{\partial x^k} - i\tilde{\Phi}_0 + \frac{i}{\left(-g^{00}\right)}\tilde{\gamma}^0\tilde{\gamma}^k\tilde{\Phi}_k$$
$$-\frac{i}{4\left(-g^{00}\right)}\tilde{\gamma}^0\tilde{\gamma}^k\frac{\partial \ln\left(gg^{00}\right)}{\partial x^k} + \frac{i}{4}\frac{\partial \ln\left(gg^{00}\right)}{\partial t}. \tag{5}$$

Recall that upper tildes denote the quantities belonging to the system of tetrads in the Schwinger gauge.

In case of interaction with a gravitational field, the covariant derivative $\nabla_\alpha \psi$ is calculated as

$$\nabla_\alpha \psi = \frac{\partial \psi}{\partial x^\alpha} + \Phi_\alpha \psi. \tag{6}$$

As we know, in the presence of the electromagnetic field, the quantity $\nabla_\alpha \psi$ transforms as follows[2]:

$$\nabla_\alpha \psi = \frac{\partial \psi}{\partial x^\alpha} + \Phi_\alpha \psi + ieA_\alpha \psi. \tag{7}$$

Here, $A_\alpha$ is the vector potential of the electromagnetic field. The comparison of (6) and (7) indicates directly, what steps should be taken to incorporate the effects of electromagnetic interaction. In particular, in formulas of [1] the following replacement should be made:

$$\Phi_\alpha \;\rightarrow\; \Phi_\alpha + ieA_\alpha. \tag{8}$$

By doing so, instead of the algorithm for constructing the generalized Hamiltonian $H_\eta$ in the $\eta$ representation from [1] we arrive at the following modified algorithm:

1) For a gravitational field described by the metric $g_{\alpha\beta}(x)$, we find a system of tetrads $\{\tilde{H}^\alpha_\mu(x)\}$ satisfying the Schwinger gauge. Recall that components of the tetrads $\tilde{H}^0_{\underline{0}}, \tilde{H}^k_{\underline{0}}$ in this gauge correlate with components of the tensor $g^{\alpha\beta}(x)$ as follows:

$$\tilde{H}^0_{\underline{0}} = \sqrt{-g^{00}}; \quad \tilde{H}^k_{\underline{0}} = -\frac{g^{0k}}{\sqrt{-g^{00}}}. \tag{9}$$

Components of $\tilde{H}^0_{\underline{k}}$ are identically equal to zero,

$$\tilde{H}^0_{\underline{k}} = 0. \tag{10}$$

In order to find $\tilde{H}^n_{\underline{m}}$, we introduce a tensor $f^{mn}$ with components

$$f^{mn} = g^{mn} - \frac{g^{0m}g^{0n}}{g^{00}}. \tag{11}$$

The tensor $f^{mn}$ satisfies the condition

---

[2] We skip the discussion of the rule for constructing covariant derivatives to account for the interaction between spin ½ particles and gauge fields. This issue is discussed in numerous monographs – see e.g. [3].

$$f^{mn}g_{nk} = \delta_k^m. \tag{12}$$

As $\tilde{H}_{\underline{m}}^n$ we can use any triplet of three-dimensional vectors satisfying the relationship

$$\tilde{H}_{\underline{k}}^m \tilde{H}_{\underline{k}}^n = f^{mn}. \tag{13}$$

2) We write a general expression for the Hamiltonian $\tilde{H}$.

$$\tilde{H} = -\frac{im}{(-g^{00})}\tilde{\gamma}^0 + \frac{i}{(-g^{00})}\tilde{\gamma}^0\tilde{\gamma}^k \frac{\partial}{\partial x^k} - i(\tilde{\Phi}_0 + ieA_0) + \frac{i}{(-g^{00})}\tilde{\gamma}^0\tilde{\gamma}^k (\tilde{\Phi}_k + ieA_k). \tag{14}$$

Here:

$$\tilde{\gamma}^\alpha = \tilde{H}_{\underline{\beta}}^\alpha \gamma^{\underline{\beta}}, \tag{15}$$

$$\tilde{\Phi}_\alpha = -\frac{1}{4}\tilde{H}_{\underline{\mu}}^\varepsilon \tilde{H}_{\nu\underline{\varepsilon};\alpha}\tilde{S}^{\mu\nu}. \tag{16}$$

3) The Hamiltonian $H_\eta$ is calculated as

$$H_\eta = \tilde{\eta}\tilde{H}\tilde{\eta}^{-1} + i\tilde{\eta}\frac{\partial \tilde{\eta}^{-1}}{\partial t}, \tag{17}$$

where the operator $\tilde{\eta}$ is given by the relationship [3]

$$\tilde{\eta} = (gg^{00})^{1/4}. \tag{18}$$

Expressions (17), (18) define the operator $H_\eta$, being the sought Hermitian Hamiltonian in the $\eta$ representation.

Thus,

$$H_\eta = -\frac{im}{(-g^{00})}\tilde{\gamma}^0 + \frac{i}{(-g^{00})}\tilde{\gamma}^0\tilde{\gamma}^k \frac{\partial}{\partial x^k} - i(\tilde{\Phi}_0 + ieA_0) + \frac{i}{(-g^{00})}\tilde{\gamma}^0\tilde{\gamma}^k (\tilde{\Phi}_k + ieA_k)$$
$$-\frac{i}{4(-g^{00})}\tilde{\gamma}^0\tilde{\gamma}^k \frac{\partial \ln(gg^{00})}{\partial x^k} + \frac{i}{4}\frac{\partial \ln(gg^{00})}{\partial t}. \tag{19}$$

The rules of the above modified algorithm can be used to find Hamiltonians in the $\eta$ representation for any stationary and non-stationary metrics and arbitrary electromagnetic fields.

The Hamiltonian in the $\eta$ representation for a Dirac particle interacting with an electromagnetic field in the spatially flat Friedmann model has the following form:

$$H_\eta = im\gamma_{\underline{0}} - \frac{i}{b(t)}\gamma_{\underline{0}}\gamma_{\underline{k}}\frac{\partial}{\partial x^k} + eA_0 + \frac{1}{b(t)}\gamma_{\underline{0}}\gamma_{\underline{k}}eA_k. \tag{20}$$

As we see from (20), in order to perform the analysis, we need to know expressions for the vector potential of the electromagnetic field.

At deriving for Hamiltonian (20) and expressions for electromagnetic field potentials, we further will use the following expressions corresponding to metric (1):

- Christoffel symbols:

---

[3] At use of curvilinear coordinates in Eq. (14) with record of volume element $dV = g_c dx_c^1 dx_c^2 dx_c^3$ it is necessary to use only a part of determinant connected with presence of external gravitational field $g_G = \frac{g}{g_c}$ [4] in Eq. (18) for $\tilde{\eta}$.

$$\begin{pmatrix} 0 \\ 00 \end{pmatrix} = 0; \quad \begin{pmatrix} 0 \\ 0k \end{pmatrix} = 0; \quad \begin{pmatrix} 0 \\ mn \end{pmatrix} = b\dot{b}\delta_{mn};$$
$$\begin{pmatrix} k \\ 00 \end{pmatrix} = 0; \quad \begin{pmatrix} m \\ 0n \end{pmatrix} = \frac{\dot{b}}{b}\delta_n^m; \quad \begin{pmatrix} k \\ mn \end{pmatrix} = 0. \qquad (21)$$

- Riemann tensor components:
$$R^0{}_{m0n} = \delta_{mn} b\ddot{b}$$
$$R^p{}_{mqn} = \dot{b}^2 \left[ \delta_q^p \delta_{mn} - \delta_n^p \delta_{qm} \right] \qquad (22)$$
$$R_{0m0n} = -\delta_{mn} b\ddot{b} \quad R_{m0n0} = -\delta_{mn} b\ddot{b}$$

- Ricci tensor components:
$$R_{00} = -3\frac{\ddot{b}}{b} \quad R_{mn} = \delta_{mn}\left[b\ddot{b} + 2\dot{b}^2\right];$$
$$R_0^0 = 3\frac{\ddot{b}}{b} \quad R_m^n = \delta_m^n \left[\frac{\ddot{b}}{b} + 2\frac{\dot{b}^2}{b^2}\right]; \quad R = 6\frac{\ddot{b}}{b} + 6\frac{\dot{b}^2}{b^2}. \qquad (23)$$

## 3. Solution of Coulomb potential type in the spatially flat Friedmann model[4]

### 3.1. General relationships for the Maxwell equations

Let us write general relationships for the Maxwell equations valid for any space.
Definition of the tensor $F_{\alpha\beta}$:
$$F_{\alpha\beta} = A_{\beta,\alpha} - A_{\alpha,\beta}. \qquad (24)$$
Lorenz gauge condition:
$$g^{\mu\nu} A_{\mu;\nu} = 0. \qquad (25)$$
This condition can be expressed as
$$g^{\mu\nu} A_{\mu,\nu} - g^{\mu\nu} \begin{pmatrix} \varepsilon \\ \mu\nu \end{pmatrix} A_\varepsilon = 0. \qquad (26)$$
Maxwell equations in the general form are written as:
$$F_{\alpha\mu;\nu} g^{\mu\nu} = 4\pi j_\alpha. \qquad (27)$$
For a continuously distributed charged substance, the vector of current density $j^\alpha$ is defined by the relationship (see e.g. [5])
$$j^\alpha = \frac{\rho}{\sqrt{-g_{00}}} \frac{dx^\alpha}{dx^0}. \qquad (28)$$
For point charges:

---
[4] The similar problem for Fermi normal coordinates is considered previously by L.Parker in [4].

$$j^\alpha = \sum_N \frac{e_N}{\sqrt{-g}} \delta(\mathbf{r}-\mathbf{r}_N) \frac{dx^\alpha}{dx^0}. \tag{29}$$

Let us calculate the left side of relationship (27).

$$F_{\alpha\mu;\nu} g^{\mu\nu} = \left(A_{\mu;\alpha;\nu} - A_{\alpha;\mu;\nu}\right) g^{\mu\nu} =$$
$$= \left(\left(A_{\mu;\alpha;\nu} - A_{\mu;\nu;\alpha}\right) + A_{\mu;\nu;\alpha} - A_{\alpha;\mu;\nu}\right) g^{\mu\nu} = \tag{30}$$
$$= \left(A_\varepsilon R^\varepsilon{}_{\mu\alpha\nu} + A_{\mu;\nu;\alpha} - A_{\alpha;\mu;\nu}\right) g^{\mu\nu} = A_\varepsilon R^\varepsilon{}_\alpha + A_{\mu;\nu;\alpha} g^{\mu\nu} - A_{\alpha;\mu;\nu} g^{\mu\nu}.$$

Using (25), we obtain:

$$F_{\alpha\mu;\nu} g^{\mu\nu} = A_\varepsilon R^\varepsilon{}_\alpha - A_{\alpha;\mu;\nu} g^{\mu\nu} =$$
$$= R_{\alpha\mu} A_\nu g^{\mu\nu} - g^{\mu\nu} A_{\alpha,\mu,\nu} + g^{\mu\nu} \begin{pmatrix}\lambda\\\mu\alpha\end{pmatrix}_{,\nu} A_\lambda + g^{\mu\nu}\begin{pmatrix}\varepsilon\\\mu\alpha\end{pmatrix} A_{\varepsilon,\nu} \tag{31}$$
$$+ g^{\mu\nu}\left[\begin{pmatrix}\varepsilon\\\nu\alpha\end{pmatrix} A_{\varepsilon,\mu} + \begin{pmatrix}\varepsilon\\\nu\mu\end{pmatrix} A_{\alpha,\varepsilon}\right] - g^{\mu\nu}\begin{pmatrix}\varepsilon\\\nu\alpha\end{pmatrix}\begin{pmatrix}\lambda\\\varepsilon\mu\end{pmatrix} A_\lambda - g^{\mu\nu}\begin{pmatrix}\varepsilon\\\nu\mu\end{pmatrix}\begin{pmatrix}\lambda\\\varepsilon\alpha\end{pmatrix} A_\lambda.$$

By rearranging the terms on the right side of (31), we can write the Maxwell equations in the following way:

$$R_{\alpha\mu} A_\nu g^{\mu\nu} - g^{\mu\nu} A_{\alpha,\mu,\nu} + g^{\mu\nu}\left[2\begin{pmatrix}\varepsilon\\\nu\alpha\end{pmatrix} A_{\varepsilon,\mu} + \begin{pmatrix}\varepsilon\\\nu\mu\end{pmatrix} A_{\alpha,\varepsilon}\right]$$
$$+ g^{\mu\nu}\left\{\begin{pmatrix}\lambda\\\alpha\mu\end{pmatrix}_{,\nu} - \begin{pmatrix}\varepsilon\\\alpha\mu\end{pmatrix}\begin{pmatrix}\lambda\\\varepsilon\nu\end{pmatrix} - \begin{pmatrix}\varepsilon\\\nu\mu\end{pmatrix}\begin{pmatrix}\lambda\\\varepsilon\alpha\end{pmatrix}\right\} A_\lambda = 4\pi j_\alpha. \tag{32}$$

Maxwell equations (32) are valid for any Riemann space written with using of gauge condition (25).

### 3.2. VECTOR POTENTIAL FOR THE SPATIALLY FLAT FRIEDMANN MODEL

In this section, Maxwell equations (32) with gauge condition (25) are solved for a point electric charge located in a space with metric (1). A similar problem in the Minkowski space is known to have a solution in the form of a Coulomb field of a point charge.

We will solve the problem using the methods of the perturbation theory. The smallness parameter is assumed to correlate with the time derivative; a more specific definition of the dimensionless smallness parameter will be given below. The arrangement of the orders of smallness, which we will use, should be confirmed by their consistency with the solved equations.

At this stage we assume that the following arrangement of the orders of smallness holds for components of the four-vector potential:

$$\left.\begin{aligned} A_0 &= A_0^{(0)} + A_0^{(2)} + ..., \\ A_k &= A_k^{(1)} + A_k^{(3)} + .... \end{aligned}\right\} \tag{33}$$

We proceed to solving Maxwell equation (32) using the methods of the perturbation theory. We can write these equations at $\alpha = 0$ in the following way:

$$R_{0\mu}A_\nu g^{\mu\nu} - g^{\mu\nu}A_{0,\mu,\nu} + g^{\mu\nu}\left[2\binom{\varepsilon}{\nu 0}A_{\varepsilon,\mu} + \binom{\varepsilon}{\nu\mu}A_{0,\varepsilon}\right]$$
$$+g^{\mu\nu}\left\{\binom{\lambda}{0\mu}_{,\nu} - \binom{\varepsilon}{0\mu}\binom{\lambda}{\varepsilon\nu} - \binom{\varepsilon}{\nu\mu}\binom{\lambda}{\varepsilon 0}\right\}A_\lambda = 4\pi j_0. \tag{34}$$

Then, in equation (34) we isolate the terms of the zero and second orders of smallness using formulas (1), (21)-(23) for the metrics, Christoffel symbols and Ricci tensor.

$$3\frac{\ddot{b}}{b}A_0^{(0)} + \ddot{A}_0^{(0)} - \frac{1}{b^2}\Delta A_0^{(0)} - \frac{1}{b^2}\Delta A_0^{(2)} + 2\frac{1}{b^2}\frac{\dot{b}}{b}A_{k,k}^{(1)} + \frac{3}{b^2}b\dot{b}\dot{A}_0^{(0)} - 3\frac{\dot{b}^2}{b^2}A_0^{(0)} = 4\pi j_0. \tag{35}$$

Here, the Laplacian $\Delta$ is understood to be the sum of the second derivative in coordinates $x, y, z$ used in (1), i.e.

$$\Delta = \frac{\partial^2}{\partial x^2} + \frac{\partial^2}{\partial y^2} + \frac{\partial^2}{\partial z^2}. \tag{36}$$

From (35) in the zero order of smallness we obtain:

$$\frac{1}{b^2}\Delta A_0^{(0)} = -4\pi j_0. \tag{37}$$

Following (29), for positive charge $Ze$,

$$j_0 = -\frac{Ze}{\sqrt{-g}}\delta(\mathbf{r}) = -\frac{Ze}{b^3}\delta(x)\delta(y)\delta(z), \tag{38}$$

and equation (37) has the form of

$$\frac{1}{b^2}\frac{1}{r^2}\frac{\partial}{\partial r}\left(r^2\frac{\partial A_0^{(0)}}{\partial r}\right) = 4\pi\frac{Ze}{b^3}\delta(r). \tag{39}$$

The solution of (39) is given by

$$A_0^{(0)} = -\frac{Ze}{br}. \tag{40}$$

In such a form, the expression for $A_0^{(0)}$ is added up to the usual expression for scalar potential, when it is written for the locally flat Minkowski space $\left(\varphi = \frac{Ze}{r}\right)$.

Further, in equation (32), we assume that $\alpha = k$, $j_k = 0$. This equation is written in the first order of smallness using expressions (1), (21)-(23) for the metric, Christoffel symbols and Ricci tensor. We obtain:

$$\Delta A_k^{(1)} = 2b\dot{b}A_{0,k}^{(0)}. \tag{41}$$

Substituting expression (40) for $A_0^{(0)}$ into (41) gives

$$\Delta A_k^{(1)} = 2\dot{b}\frac{Zex_k}{r^3}. \tag{42}$$

It follows from (42) that

$$A_k^{(1)} = -Ze\dot{b}\frac{x_k}{r}. \tag{43}$$

We continue solving equation (35) in the second order of smallness:

$$3\frac{\ddot{b}}{b}A_0^{(0)} + \ddot{A}_0^{(0)} - \frac{1}{b^2}\Delta A_0^{(2)} + 2\frac{1}{b^2}\frac{\dot{b}}{b}A_{k,k}^{(1)} + \frac{3}{b^2}b\dot{b}\dot{A}_0^{(0)} - 3\frac{\dot{b}^2}{b^2}A_0^{(0)} = 0. \tag{44}$$

Substituting the expression for $A_{k,k}^{(1)}$ from (43) gives:

$$\Delta A_0^{(2)} = -2\frac{Ze}{r}\ddot{b}. \tag{45}$$

Hence,
$$A_0^{(2)} = -Ze\ddot{b}r \tag{46}$$

Thus, final expressions for the four-vector potential components are written as:
$$A_0 = -\frac{Ze}{br} - Ze\ddot{b}r + ..., \tag{47}$$

$$A_k = -Ze\dot{b}\frac{x_k}{r} + .... \tag{48}$$

Expressions (47), (48) satisfy gauge condition (25), that shows that the method of finding solutions to the Maxwell equation is self-consistent.

Conservation law of electric charge is fulfilled for solutions (47), (48) in every orders of smallness.

Formulas (47), (48) confirm the assumptions about the arrangement of the orders of smallness made at the beginning of the discussion.

As we see, in the spatially flat Friedmann model, the point charge problem cannot be solved in principle without accounting for the $A_k$ components. In this respect, the problem in the Friedmann model differs qualitatively from the similar problem in the flat space-time.

It follows from (47), (48) that potential energy $U$ in the Hamiltonian for an electron in a hydrogen-like atom is given by
$$U = -\frac{Ze^2}{br} - Ze^2\ddot{b}r. \tag{49}$$

The Hamiltonian also includes interactions with the $A_k$ electromagnetic field components.

The second summand in the right part of (49) has the second order of smallness in relation to the first one.

Let us estimate the magnitude of the smallness parameter used in the expansions.

By definition, $\dot{b} = Hb$. In the order of magnitude, $\ddot{b} \sim H^2 b$, where $H$ is the Hubble constant. At present, $H_0 = 2,4 \cdot 10^{-18} \frac{1}{\sec}$.

Potential energy (49) can be written as
$$U \approx -\frac{Ze^2}{br} - \frac{H^2}{c^2} Ze^2 br \tag{50}$$

In expression (50), $c$ is the speed of light; in the foregoing, we used the system of units $\hbar = c = 1$.

For atomic distances of $r \approx 10^{-8}$ cm, the ratio of the second summand to the first one in (50) is equal to
$$\frac{(2,4 \cdot 10^{-18})^2}{(3 \cdot 10^{10})^2} b^2 (10^{-8})^2 \sim 10^{-72} \cdot b^2 \ll 1. \tag{51}$$

The inequality (51) with big reserve is valid at the change $b(t)$ from 1 at present to $10^{-3}$ in the cosmological past (recombination epoch of atomic system, see, e.g., [6]).

As we see from the left side of (51), the smallness parameter for atomic systems is the ratio of atomic dimensions $a \approx 10^{-8} cm$ to the size of the universe $\frac{c}{H_0} \approx 10^{28} cm$,
$$\lambda \sim \frac{aH_0}{c} \sim 10^{-36} \tag{52}$$

# 4. COUPLED ATOMIC SYSTEMS AND FORCES OF INTERACTION BETWEEN ELEMENTARY PARTICLES IN THE SPATIALLY FLAT FRIEDMANN MODEL

## 4.1. HYDROGEN-LIKE ATOM IN THE EXPANDING UNIVERSE

Let us consider the motion of an electron in the Coulomb field of a nucleus with charge $Ze$ in the spatially flat Friedmann model. With Hamiltonian (20) and expression (47) for the scalar potential $A_0$ with the leading first summand, the Dirac equation can be written as

$$i\hbar \frac{\partial \psi}{\partial t} = \left( c\boldsymbol{\alpha} \frac{\mathbf{p}}{b(t)} + \beta mc^2 - \frac{Ze^2}{b(t)r} - \frac{Ze\dot{b}}{b(t)r}\boldsymbol{\alpha}\mathbf{r} + ... \right)\psi . \qquad (53)$$

Equation (53) uses the metric with the signature $\eta_{\alpha\beta} = \mathrm{diag}[1,-1,-1,-1]$ with four-dimensional matrices in the Dirac-Pauli representation $\left(\alpha^k = \gamma^0\gamma^k,\ \beta = -i\gamma^0\right); p^i = -i\hbar\frac{\partial}{\partial x^i}$.

The function $b(t)$ in equation (53) implies that characteristics of atomic systems in principle can vary on the cosmological time scale. In the principal order of the perturbation theory, at each point of cosmological time one can solve the problem of finding wave functions and energy spectrum in the quasi-stationary approximation. In this approximation, the quantity $b(t)$ is constant, and metric (1) reduces to the metric of the Minkowski space by coordinate transformation and the term $\frac{Ze\dot{b}}{b(t)r}\boldsymbol{\alpha}\mathbf{r}$ in (53) can be omitted because of the parameter smallness (52).

So, we decide the task of finding of wave functions and energy spectrum in each point of cosmological time in quasi-stationary approximation. At solution of equation (53) can be used standard procedure of finding of power levels [8], [9].

At the solution of (53) in quasi-stationary approximation it is easy to prove that energy levels for Dirac equation (52) do not depend on the quantity $b(t)$ and that they coincide with the standard solution [8], [9].

$$E_n = mc^2 \left[ 1 + \left((Z\alpha) \bigg/ \left( n - \left(j+\frac{1}{2}\right) + \sqrt{\left(j+\frac{1}{2}\right)^2 - Z^2\alpha^2} \right)\right)^2 \right]^{-\frac{1}{2}}. \qquad (54)$$

Here, $\alpha = \frac{e^2}{\hbar c}$ is the fine-structure constant, $n, j$ are quantum numbers. Naturally, the same conclusion is also drawn for the energy levels in the non-relativistic Schrödinger equation.

The solutions to equation (52) in the quasi-stationary approximation have the following features. Energy spectrum (53) is invariant with respect to the identical variation of coordinate values of momentum $\mathbf{p}$ and coordinate potential energy $U = -\frac{Ze^2}{r}$. Solution (53) is identical for both extreme low and extreme high values of $p^i_{phys} = \frac{p^i}{b(t)}$, $U_{phys} = -\frac{Ze^2}{b(t)r}$. Ionization potentials remain constant over the entire range of variation of $b(t)$ (for the hydrogen atom, $I = 13.6\,eV$). Of course, for very small values of $p^i_{phys}$ and $U_{phys}$, atomic states will decay due to quantum fluctuations in accordance with the Heisenberg uncertainty principles.

Thus, the structure of atomic levels does not vary with universe expansion. This implies that our results confirm the standard interpretation of the mechanism of redshift as a shift associated only with the growth of wavelength of photons as they propagate through the universe.

Also, note that in the quasi-stationary approximation the effect of spectral line splitting is negligibly small due to the influence of the gravitational field connected with universe expansion.

Previously the analogous conclusion for the spatially flat universe model was made by J.Audretsch and G. Schäfer in papers [10]. But the self-conjugacy of Dirac Hamiltonian reached with the aid of a non-unitary transformation did not seem substantiated enough. In addition, in these papers the vector potential of electromagnetic field is assumed at a glance in form of the dominant term (40) without corrections (43), (46). These corrections, in spite of its smallness (see (52)), are necessary for fulfillment of gauge condition (25) and self-consistency method of solution of the Maxwell equations.

The wave functions obtained by solution of (53) in quasi-stationary approximation do not change during universe expansion, if one can use in coordinate presentation not a radial variable $r$, but a quantity $r_{phys} = br$.

For Equation (53) a radius of the Bohr orbit is defined from the following relationship

$$a_{phys} = b(t)a, \text{ where } a = \frac{\hbar^2}{me^2} \tag{55}$$

The relationship (55) shows that the atomic systems in the cosmological past had the smaller physical dimensions then at present $t = t_0$. If $b(t_0) = 1$, the physical radius of the Bohr orbit was $a \simeq 10^{-11}$ cm for formation time of atomic system (recombination epoch, $b \approx 10^{-3}$).

## 4.2. ENERGY OF INTERACTION BETWEEN ELEMENTARY PARTICLES IN THE EXPANDING UNIVERSE

Primarily, let us consider electromagnetic coupling. An interaction energy, according to (49), consists of two terms: the Coulomb energy $-\frac{Ze^2}{br}$ and energy $-Ze^2\ddot{b}r$. The second term is negligibly small (see (52)) both for short and reasonable large (cosmological) distances.

The Coulomb energy and energy connected with presence of vector-potential $A^k$, according to the Hamiltonian (20), are changed with cosmological time $\sim \frac{1}{b(t)}$. In the past an energy of electromagnetic interaction was high, but in the cosmological future it will be infinitesimal.

Let us discuss the question about force influenced on a charge $e$, which is placed in electromagnetic field with potentials (47), (48). Calculation of electric field intensity $(E_k)_{phys}$ and magnetic induction $(B_k)_{phys}$ up to the second order of smallness gives

$$(E_k)_{phys} = \frac{1}{b}E_k = -\frac{1}{b}F_{0k} = \frac{Ze}{b^2} \cdot \frac{x_k}{r^3}; \quad (B_k)_{phys} = \frac{1}{2b}\varepsilon_{kmn}F_{mn} = 0 \tag{56}$$

Such form for vector $(E_k)_{phys}$ in each order of smallness automatically guaranties a fulfilment of a relation

$$\frac{1}{b^2}E_{k,k} = 4\pi j^0 \tag{57}$$

and thereby a fulfilment of conservation law of electric charge. As for the force influenced on charge $e$, it is defined by the Lorentz equation, which in our case is added up to the following expression

$$\left(F_{e.m.}\right)_{phys} = \frac{e}{b}E_k = -\frac{e}{b}F_{0k} = \frac{Ze^2}{b^2} \cdot \frac{x_k}{r^3} \tag{58}$$

As we see from (58), the force decreases as universe expansion.

Let us now address strong interactions and, in particular, the phenomenological potential $QCD$ modified by analogy with electromagnetic potentials for using it in the spatially flat expanding universe $\left(r \to r_{phys} = b(t)r\right)$:

$$F_{QCD} = const; \quad V_{phys} = Kb(t)r \tag{59}$$

where $K$ is the force constant of strong interactions.

One can see from (59) that in the early universe with $b(t) \ll 1$ (if at present $b(t_0) = 1$) the potential $V_{phys}$ is low, and it provides almost no coupling of quarks and gluons. As $b(t)$ grows, the quark-gluon coupling becomes stronger. One can say that the expansion of the universe reinforces the confinement of quarks inside hadrons and mesons.

The relationships (55), (59) show that hadrons, mesons and atomic systems had the denser structure with small physical dimensions in the formation time. The quark-gluon structures and atoms become less dense and increase its characteristic physical dimensions as universe expansion.

## 5. Conclusions

This study develops the results of [1] related to the form of the Hamiltonian for spin ½ particles in an arbitrary not only gravitational but and electromagnetic field. The generalized Hamiltonian in the $\eta$-representation is given by formula (19).

In Sec. 3 we obtained solutions to the Maxwell equations in the spatially flat Friedmann model. The solutions for component of vector-potential $A_0$, $A_k$ have the form of (47), (48). It follows from them that in the spatially flat Friedmann model, the point charge problem cannot be solved in principle without accounting for the $A_k$ components. In this respect, the problem in the Friedmann model differs qualitatively from the similar problem in the flat space-time. Remind, that in standard cosmological model the spatially flat Friedmann model is considered as model, which explains in the best way the observed data, including universe expansion with acceleration.

On the basis of the Hamiltonian in the $\eta$-representation (19) for the Dirac particle which is written for the concerned cosmological model, in Sec. 4 we considered atomic systems and electromagnetic forces of interaction for the spatially flat expansion of the universe. It is shown (and it is the central result of this study) that the system of atomic levels of hydrogen-like atoms does not vary with cosmological time. And, on the contrary, the forces of interaction between elementary particles vary as the universe expands.

Thus, spectral lines of atoms in the spatially flat expanding universe are identical at different points of cosmological time. This statement validated in this study within of the general relativity theory and the theory of self-conjugated Hamiltonians in the $\eta$-representation is a strict result, which confirms the concepts of the standard cosmological model about the mechanism of redshift of atomic spectra.

Of cause, the above discussions imply applicability of the spatially flat Friedmann model for considered periods of the cosmological times.

## ACKNOWLEDGEMENTS

The authors would like to thank Prof. P.Fiziev, B.P.Kosyakov for the useful discussions, advices and criticism.